# Molecular and Ionic Dipole Effects on the Electronic Properties of Silicon Grafted Alkylamine Monolayers

Alina Gankin[1,2], Ruthy Sfez[1,3], Evgeniy Mervinetsky[1,2], Jörg Buchwald[4], Arezoo Dianat[4], Leonardo Medrano Sandonas[4,5], Rafael Gutierrez[*,4], Gianaurelio Cuniberti[4,6] and Shlomo Yitzchaik[*,1,2]

[1] Institute of Chemistry, The Hebrew University of Jerusalem, Safra Campus, Givat Ram, Jerusalem 91904, Israel.

[2] Center for Nanoscience and Nanotechnology, The Hebrew University of Jerusalem, Jerusalem 91904, Israel.

[3] Azrieli, College of Engineering, Jerusalem, Israel.

[4] Institute for Materials Science and Max Bergmann Center of Biomaterials, TU Dresden, 01069 Dresden, Germany.

[5] Max Planck Institute for the Physics of Complex Systems, 01187 Dresden, Germany.

[6] Dresden Center for Computational Materials Science, TU Dresden, 01062 Dresden, Germany.

E-mail: Shlomo.yitzchaik@mail.huji.ac.il

**ABSTRACT:** The need to control the electronic properties of hydroxyl-terminated substrates such as Si/SiO2 or Indium-Tin oxide (ITO) is of great importance in both electronics and optoelectronics device applications. Specifically, the relevant electronic properties are the work function (WF) and the electron affinity (EA) of the substrate, which can be tailored using monolayers of polar or charged molecules constructed via covalent or electrostatic self-assembly. In this work, we demonstrate the tailoring and characterization of the electronic properties of Si substrate by creating molecular and ionic surface dipoles. The Si substrates were modified by adsorption of alkylamine terminated moiety followed by quaternization of the amine moiety to ammonium using haloacids thus exchanging the labile halide anion from Cl- to Br- and finally to I-. This exchange induced a change of the work function due to different ionic dipoles formation between the positively charged ammonium moiety and the labile halide counter-anion. All substrates were characterized by AFM, XPS, spectroscopic ellipsometry and contact angle measurements. Contact potential difference (CPD) method was used for measuring the change in the electronic properties, such as the work function (WF), of the modified Si substrate upon ionic and molecular dipole formation. The experimental results obtained by CPD were supported theoretically by DFT calculations of the ionic and molecular dipoles, giving a good fit with the obtained results and indicating the polarizability parameter as the most influencing one on the work function.

## Introduction

The need to design and tailor the electronic properties of semiconductors and metal oxides by self-assembled monolayers (SAM)[1,2] of organic molecules is growing rapidly in organic electronics[3] and optoelectronics devices.[4,5] Major efforts were made in order to understand and evaluate the parameters contributing to the change in substrate' electronic properties, such as work function (WF),[6] electron affinity (EA)[7] and band bending (BB),[7] both theoretically and experimentally. Indeed, recent reviews evaluate experimentally[8] and theoretically[9] the influence of various parameters such as molecular density[10], tilt angle, polarizability and more, on the dielectric constant of a layer, confirming that the total effect is a combination of molecular and structural parameters, as was shown experimentally numerous times.[11] It is well known that polar organic SAMs can tune the electronic properties of the semiconductors[12] or metals substrates by shifting the surface potential,[13–15] the carrier density,[15,16] the electron affinity,[17–19] and work function.[19–21] The dielectric constant of the layer is directly related to the change in the EA parameter of a modified substrate, or to the change in the WF, in case of metal or highly doped substrates (which don't have any change in BB contribution) by Helmholtz equation (eq. 1):

$$\Delta \Phi_s = \frac{N\mu}{\varepsilon \varepsilon_0} cos\theta \quad (1)$$

where N is the dipole density (in molecule/ m²), µ is the dipole moment (in Debye, 1 D = 3.34 × 10⁻³⁰ C·m), θ is the average tilt angle of the dipole with respect to the surface normal, Ɛ is the effective dielectric constant of the molecular film, and Ɛ0 is the permittivity of vacuum; here $\Delta\phi_s$ is expressed in units of volts.

Experimentally, those electronic properties can be measured by Kelvin probe[22–24], contact potential difference (CPD) measurements. Three parameters contribute to the work function of a semiconductor: the BB, the EA and the difference between the Fermi level and conductive band in the bulk. Since the difference between the Fermi level and the conductive band cannot be determinate easily[25], only relative values are used, giving for n-type Si the fallowing equation (eq 2):

$$\Delta CPD = \Delta \Phi = \Delta EA + \Delta BB \qquad (2)$$

These parameters need to be measured separately. Illumination of the sample[26] or using highly doped Si produces band flattening[27], hence $\Delta BB \sim 0$ and hence $\Delta CPD \sim \Delta EA$.

In order to estimate the influence of a reactive, usually polar, layer on a substrate, only relative parameters ($\Delta EA$, $\Delta BB$, $\Delta WF$) are calculated, by subtracting the reference value of the bare substrate from the modified one. The fact that the work function of a substrate, can be easily tuned by varying the adsorbed, usually polar, organic monolayer was used a lot to tailor desired electronic properties of such substrates.[28] The modification of the substrate can be obtained either covalently or electrostatically.[29] Basically, two main synthetic methods were applied for the layer formation: adsorption of covalently attached molecules with specific head groups[30] and tailored tail group[31] by molecular self-assembly,[32,33] or by polyelectrolyte assembly of charged polymers[34] with a electrostatic interactions.

In previous works we have shown that the WF and EA of a Si/SiO2 substrate can be tuned by a monolayer of polar chloro-[35] or alkoxy-silanes[36] monolayers with various tail groups. It was shown that polar and polarizable molecules with dipoles pointing toward the substrate, decrease the work function of n-type substrate while dipoles pointing away from the surface, increase the work function. It was also shown that while using propylhalide (PrX) tail groups (i.e. PrI, PrBr, PrCl) bearing trichlorosilane head groups, the ΔEA parameter decreased, with decrease of the dipole. Namely, for halogen based terminated groups such as I, Br and Cl tail groups, with molecular dipoles of 1.47 D, 1.60 D and 1.63 D, the ΔEA parameter observed was of 0.21, 0.28 and 0.38 eV respectively.[35]

In this work, we demonstrate both theoretically and experimentally. The WF tunability of Si/SiO2, via both directed ionic and molecular dipole changes. Namely, Si substrates were modified by 3-Aminopropyltriethoxysilane (APTES), which has an amine terminated tail groups. The ionic-dipole was achieved by a gradual protonation of the amine groups to ammonium groups using pH decreasing. Additionally, a change in ionic dipole was achieved by a changing the halide counter-ion of a positively charged alkylamonium tail group of covalently anchored monolayer, from Cl- to Br- and I-. Both molecular and ionic dipole changes on Si/SiO2 substrate have shown a major change in the work function.

## Experimental

CPD measurements were conducted with a commercial instrument, Kelvin probe S (DeltaPhi Besocke, Jülich, Germany), with a vibrating gold electrode (work function 5.1 eV) in a home-built Faraday cage under Ar (argon) atmosphere. Variable Angle Spectroscopic Ellipsometry (VASE) measurements were carried out with ellipsometer VB-400, (Woollam Co.) at the Brewster angle for silicon 75°. Contact angle (CA) measurements were carried out by with TDW using Attension device goniometer (Theta Life, KSV Instrument, Biolin Scientific). X-Ray Photoelectron Spectroscopy (XPS) spectra were collected at ultrahigh vacuum (2.5 x 10$^{-10}$ Torr) on a 5600 Multi-Technique (AES/XPS) system (PHI) using an X-ray source of Al Kα (1486.6 eV). Atomic force microscopy (AFM) measurements were carried out with a Nanoscope IV (Veeco Dimension 3100) in tapping mode using a tapping etched silicon probe (TESP, DI) with a 30 N/m force constant.

**Chemicals**: 3-aminopropyltriethoxysilane (APTES, 99%, Aldrich) was distilled before use. HCl was purchased from Sigma Aldrich. Ethanol HPLC graded, toluene, $H_2SO_4$, $H_2O_2$, HBr were purchased from Merck, HI was purchased from Across organics, acetonitrile and 2-propanol were purchased from J.T. Baker and used with no further purification.

**Surface modification:** Highly doped n-Si <100> (Si R< 0.003 Ω/cm) substrates were cleaned in aqueous detergent, washed with triple distilled water (TDW), dipped in piranha solution ($H_2O_2$/$H_2SO_4$ concentrated, 3:7 v/v) for 15 min The substrates were then rinsed with TDW and sonicated in $NH_3$/$H_2O$/$H_2O_2$ (1:1:5 v/v/v) solution for 30 min at 60°C.[37] The substrates were washed with TDW, pure acetone and then dried under a stream of nitrogen. Freshly cleaned and activated substrates of Si/SiO$_2$ were immersed in 0.2% (v/v) APTES /ethanol solution for 20 min. After the proper immersion time, the Si/SiO$_2$ substrates were washed three times with ethanol and dried under a stream of nitrogen, followed by curing in an oven at 100 °C for 30 min. Si/SiO$_2$-PrNH$_2$ substrates were used in two different experiments: 1) immersion in HCl solution in different pH (1-5) for 5 minutes; 2) immersion in 0.1 M solution of HCl, HBr, HI for 5 min followed by drying under a stream of nitrogen.

**Theoretical modeling:** In order to rationalize the observed experimental trends in the work function of molecular and ionic dipoles, we additionally performed density functional theory based calculations. For this purpose, we used a 15.36 Å × 7.68 Å reconstructed Si(100)-(2x1) surface with six interconnected aminopropyl/propylhalide silicon dioxide molecules attached to it. All calculations were done using Quantum Espresso software package[38]
 with projector augmented-wave pseudopotentials and the Perdew-Burke-Ernzerhof (PBE) exchange-correlation functional[39,40] with non-local dispersion corrections (vdW-DF) .[41,42,43] The kinetic energy cutoff was set to 544 eV while the charge density cutoff was chosen to be ten times higher. The integration over the Brillouin zone was performed employing a 3×6×1 Monkhorst-Pack mesh.[44] The distance between adjacent slabs was set to 20 Å and the dipole correction of Bengtsson[45] was used. The atomic positions were relaxed until all force components on every atom were less than $2 \times 10^{-3}$ Ry/$a_o$ and the changes in total energy are less than $2 \times 10^{-4}$ Ry.

**Results and Discussion**

A monolayer of APTES was assembled on Si/SiO$_2$ (Figure 1) as described elsewhere.[46] AFM measurements verify smooth and continuous monolayer formation (SI, Figure S1). This step results in a layer containing both alkylamonium (pKa~10) and alkylamine functionalities. The ratio between the ammonium and amine groups grows with decreasing pH.

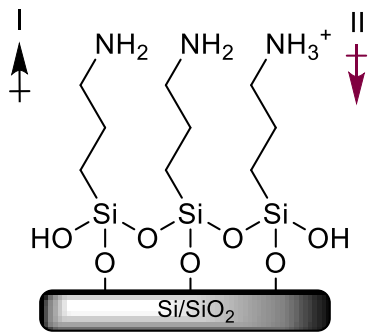

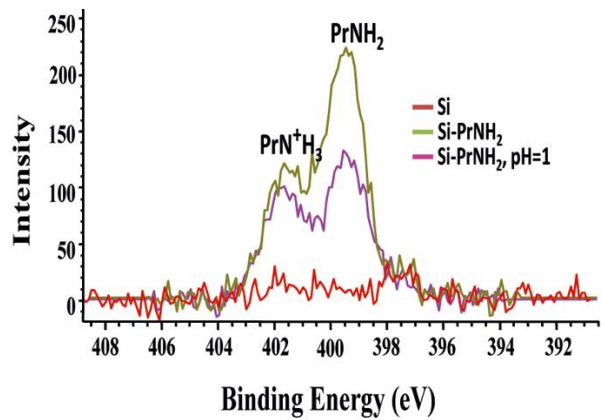

**Figure 2.** XPS of bare silicon, Si/SiO$_2$-PrNH$_2$ and Si/SiO$_2$-PrN$^+$H$_3$ after immersing in HCl at pH=1.

**Figure 1.** Schematic description of the dipole direction in PrNH$_2$ and PrN$^+$H$_3$ containing layer on Si/SiO$_2$ substrate. (I) dipole pointing away from the surface; (II) dipole pointing toward to surface.

Contact angle (CA) analysis of the obtained film showed an increase in the CA from 8° for bare and activated silicon to 55° for Si/SiO$_2$-PrNH$_2$. Ellipsometric VASE measurements at incident angle of 75° (Cauchy model with refractive index 1.45 was used for fitting) showed a thickness of 6.7 ± 0.3 Å. XPS showed two peaks of nitrogen: N [1s] 399.8 eV, 402.2 eV which corresponds to propylamine (PrNH$_2$) and propylammonium (PrN$^+$H$_3$) respectively. Following amine terminated layer formation, a more acidic environment with lower pH (pH = 1-5) was obtained by immersing the APTES modified substrates in hydrochloric acid in order to control the PrN$^+$H$_3$/PrNH$_2$ ratio in the layer. Due to terminal amine moiety protonation, as expected, an increase of the ratio was obtained by XPS analysis, from 0.55 (for PrNH$_2$ in ethanol) to 0.75 (at pH = 1; see, Fig. 2).

The change in the work function was found to be reversible, implying that after washing with water the work function returns to its original value. It is important to notice that the ammonium groups have a dipole pointing towards the surface while the amine groups have a dipole pointing away from the surface,[36] as illustrated in Figure 1. This fact can be used for work function tuning by protonation with different pH. It should be expected that higher degree of protonation inducing more dipoles pointing toward the surface should result in the decrease of the WF. However, the CPD measurements have shown an opposite behavior resulting in increasing of the WF upon the protonation as shown in Figure 3 (see also control experiments with bare Si: SI, Figure S2).

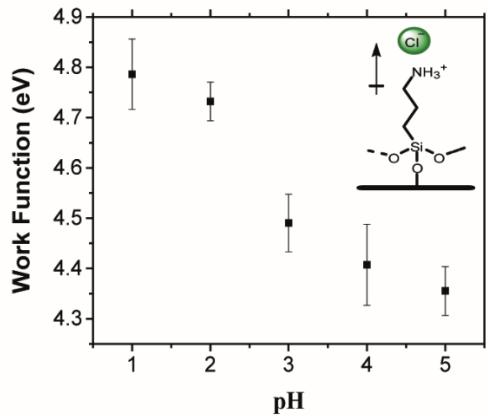

**Figure 3.** Dependence of the work function of Si/SiO$_2$ substrate with propylamine monolayer following exposure to HCl solution of different pH. The insert is schematic representation if the ionic dipole formed by Cl$^-$.

These results can be explained by the counter-ion dipole (Cl⁻), coupled to the propylammonium positively charged monolayer. The position of the Cl⁻ with respect to the PrN⁺H₃ group can create an ionic dipole with an opposite direction, which points away from the surface resulting in increasing of the WF (Figure 3). In this case the ionic dipole leads to an increase of work function from 4.24eV (Si/SiO₂-PrNH₂, ethanol) to 4.73 eV (Si/SiO₂-PrNH₂, pH=1 HCl). In order to confirm this assumption, XPS measurements were conducted in several pH values and indeed XPS result verified the quaternization reaction by the presence of Cl⁻ (Cl⁻ [2p] 198.2 eV) at pH = 1. (SI, Table S1 )

At this point, it seems of interest to explore other anions as well. A change in the counter anion was conducted by using the appropriate acids (HBr, or HI) in order to check the influence of the anions alteration on the work function of the substrates. For this purpose, some of the amine terminated substrates was immersed in hydrobromic acid, and some in hydroiodic acid. The base peak of hydrobromic acid was found in XPS measurements (Br⁻ [3d] at 0.65eV), but no indication for the I⁻ presence was obtained by XPS (due to its volatility under the x-ray radiation). CPD measurements have shown increasing of the work function with increasing of atomic radius/polarizability of the bases (Fig. 4).

It is worth noting that this changes of the work function were caused by the changes of electron affinity, and not due to the band bending (SI, Figure S3) owing to the highly doped Si used in this experiments. It is important to notice that a change in the BB occurs too (SI, Figure S3). The BB parameter is related to electron traps and surface states. In previous work it was already shown that as tail groups changing from Chlorine to Iodine in a chlorosilane monolayer, the BB increases.[47] In the present case the opposite trend is observed for labile halide anion, although due to silicon's high doping level this parameter is negligible, especially while comparing to the change in EA-ΔEA. In any case, ΔBB for the iodide is opposite to the change in EA, thus showing the important contribution of the ionic dipole to the work function.

The increasing of WF as a result of counter ions alteration, probably derives from the anions exchange that occurs between the hydroxyl ions (before exposure to acids) and the other counter anions conjugated bases (PrNH₂, pH = 1). Due to the small size of the hydroxyl anion and its capacity for hydrogen bonding, it seems energetically favorable that it lies in the ammonium plane. Upon anion exchange there is a possibility of changes in the dipole's tilt angle. There are van der Waals (vdW) interactions and hydrogen bonding between hydronium ions (from the ammonium layer) and PrN⁺H₃ in Si/SiO₂-PrNH₂ substrate. However, the interactions between conjugated bases (Br⁻, Cl⁻, I⁻) are mainly electrostatic. Therefore, interaction of PrN⁺H₃ with the hydroxyl ion can be stronger and with smaller tilt angle than the interactions with the conjugated bases. So a possible increase of the tilt angel with increasing the ionic radius (and polarizability) of conjugated bases may be considered. Such reorientation can result in a decrease of the vectorial dipole pointing towards the surface (Fig. 1).

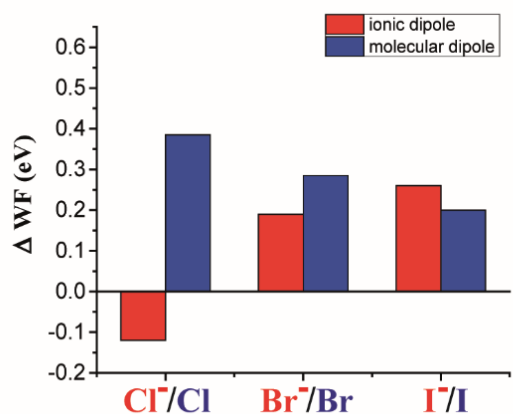

**Figure 5.** Experimental change in work function, ΔWF measurements comparison of molecular[35] and ionic dipoles on modified Si, compared to bare Si as a function of surface functionality, either ionic or covalently attached.

Figure 5 shows the difference in the work functions comparing bare Si/SiO₂ substrates for propylhalide coupling agents from a precedent work,[35] and the change in the WF in this case upon anion exchange with the halides anions. It is clear that the trend induced for the ionic dipole is opposite from the molecular dipole. Namely, the WF of modified Si with propylhalide molecules, increase while changing the halo atom from I to Cl, while the WF decrease for the ionic dipole induced by the halide anions upon changing from I⁻ to Cl⁻. Regarding the molecular dipole the change can be explained by the fact that the dipole pointing out of the surface decreases upon changing from Cl to I, thus inducing an increase in the WF. On the other hand, while ionic dipole is of importance, many parameters might influence the dipole. For example, the diameter of the anion (varying from 360, 400, 440 pm for chloride, bromide and iodide respectively [48,49]) might cause the anion to be above the ammonium moiety, thus causing a change in the dipole orientation and an increase in the WF measured. Moreover, polarizability might influence greatly the observed dipole; ergo the WF will increase upon increasing polarizability. The importance of the counter ion position on the electronic properties of a self-assembled monolayer was already observed with monolayers of polarizable molecules for nonlinear optics (NLO) applications. Translocations in anion' position where the cation is anchored covalently to the surface, induces large changes in the molecular hyperpolarizability.[47] It seems that all the parameters coincide with giving rise a larger work function for the ionic dipole with changing from smaller and less polarizable chloride ion to the large and polarizable iodide ion.

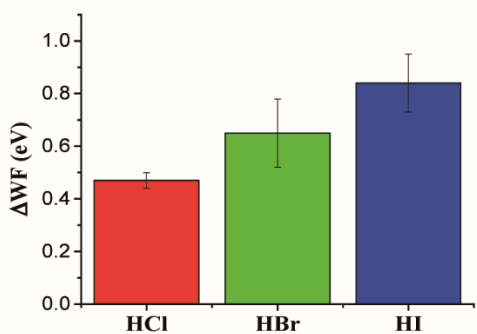

**Figure 4.** ΔWF of Si/SiO2-PrNH2 after immersing the layer in HCl, HBr, HI at pH = 1.

As previously mentioned, we carried out DFT-based calculations to gain insight into the experimental results. For this purpose, we prepared a periodic supercell with a dimerized silicon (100) slab, with six interconnected aminopropyl/propylhalide silicon dioxide molecules on top (Fig. 6). The two remaining silicon sites as well as the lower surface were passivated by hydrogen atoms. The slab thickness was chosen to be big enough for the lower surface in order to have no significant influence on the work function. While for molecular dipoles, every molecule is terminated by a halogen atom, for the ionic dipoles we have a fraction of one third $NH_2$- to $NH_3^+$-termination plus a halogen counter ion (see Fig. 1). During the relaxation, we found that some of the ions tend to sit on top of the ammonium ions, while others like to sit more within the alkylamine monolayer resulting in a different ion-counterion distances.

The WF for such a system is given by the difference between the electrostatic potential in vacuum and the Fermi level, which is typically placed in the middle of the band gap for intrinsic semiconductors (eq 3):

$$WF = -e\phi_{vac} - E_F \quad (3)$$

Using Gaussian smearing with a spreading of 0.011 Ry we calculated the Fermi energy of the slab. In a second step, we found the vacuum level of the electrostatic potential by plane averaging the electrostatic potential (eq 4)

$$\phi(z) = \frac{1}{s}\int_S \varphi(r)dxdy . \quad (4)$$

Where S is the surface area of the periodic supercell. Consequently, we found a WF value of 4.45 eV for the Si (100) surface with a siloxane monolayer on top. For this calculation, the remaining free silicon binding sites were passivated by adding H atoms.

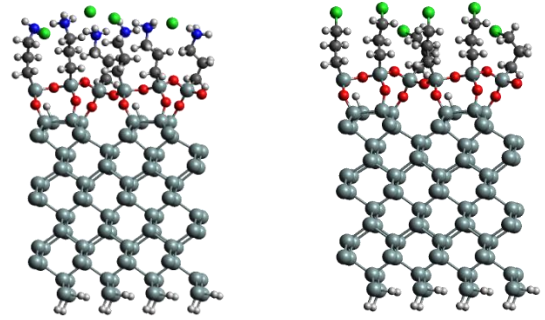

**Figure 6.** Supercell of an alkylchloride (right) and alkylamine monolayer with chloride counterions (left) on a silicon surface.

**Table 1. Calculated values of the electronic properties of the modified substrate induced by ionic dipoles**

| Ion type | WF | Dipole-moment (z-direction) | q |
|---|---|---|---|
| | [eV] | [D] | [e] |
| Cl | 5.76 | -4.35 | -0.76 ± 0.03 |
| Br | 5.913 | -4.63 | -0.74 ± 0.02 |
| I | 6.1456 | -6.17 | -0.69 ± 0.02 |

**Table 2. Calculated values of the electronic properties of the modified substrate induced by molecular dipoles**

| Ion type | WF | Dipole-moment (z-direction) | q |
|---|---|---|---|
| | [eV] | [D] | [e] |
| Cl | 5.458 | -3.21 | -0.27 ± 0.02 |
| Br | 5.402 | -3.02 | -0.15± 0.01 |
| I | 5.1 | -2.08 | 0.05± 0.02 |

Accordingly, we determined the WF also for the molecular and ionic dipoles (Tab. 1 and Tab. 2). Additionally, we also show in the table the corresponding Bader charges $q$ of the halogens, both for molecular and ionic dipoles. The errors denote the standard deviation between different halogen atoms in one supercell (cf. Fig. 6). To compare these values with the experimental, we've also plotted the difference versus bare Si in Figure 7.

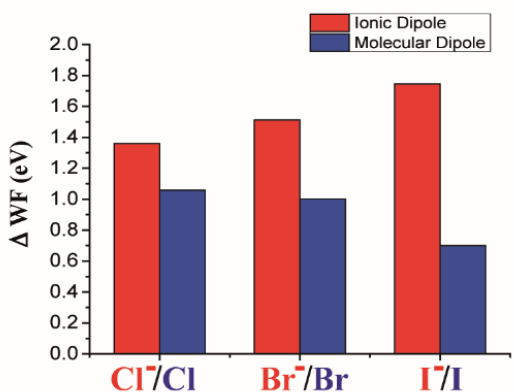

**Figure 7.** Calculated change in work function, ΔWF measurements comparison of molecular and ionic dipoles on modified Si, compared to bare Si as a function of surface functionality, either ionic or covalently attached.

Likewise, in experiments, we are able to observe similar trends: As the WF for the molecular dipoles decreases, it increases for the ionic dipoles from Cl to I. To better understand this behavior we additionally performed a Bader charge analysis.[50] In Tab. 1 and Tab. 2 we give the mean values of the Bader charges for all halogen atoms in the supercell. Therein, one sees that the negative charges decrease from Cl to I, meaning that the charge transferred from the halogens to the the surface increases, especially for the molecular - but also for the ionic dipoles. This can be explained by a weaker binding of the outer shell electrons for the heavier halogen atoms. The changes between different kinds of Halogens are thereby considerably lower for the ionic dipoles, which can be traced back to the weaker interactions of the ionic binding. Our simulations indicate that for the molecular dipoles changes in the work function are driven by charge transfer. However, in the case of ionic dipoles we find only slight changes in the corresponding Bader charges (which are of the order of the standard deviation), although the mean distances between the ammonium and the halide ions increase from 3.15 Å for Cl to up to 3.56 Å for I. Also within one halogen type, we found that greater ion-counter ion distances are associated to higher charge values, but these changes are also small (cf. error given in Tab. 1) compared to the observed differences in the binding distance, that are up to 0.5 Å. As the charges remain nearly constant for the ionic dipoles, we conclude that the main contribution to the work function is the growing ion-counter ion distance implying an increase in the dipole moment. Earlier calculations also revealed, that the observed trends in the work function depend on the stabilizing effect of the siloxane monolayer, while the attachment of single propylhalide-/amonipropylhydroxysilane molecules give different results (See SI Figure 4-Figure 7 for further details).

Changes in the absolute values are explained mainly by differences in the monolayer density and in the $NH_2/NH_3^+$ ratio comparing these results to experiments (Fig. 5).

## Conclusions

We have shown in this work an easy way of controlling and tailoring the work function of hydroxyls terminated substrates by ionic and covalent surface dipoles. $Si/SiO_2$ substrates were used for tuning the ionic dipole of amine/ammonium terminated tail group monolayer. This was done by simple neutralization reaction between propylamine moieties covalently attached to the substrate followed by immersion in the chosen halide based acid. An increase in the work function was observed upon pH decreasing for HCl. For the other haloacids, a clear trend was observed of increase in the work function while changing the counter ion from $Cl^-$ to $I^-$. Calculations have supported the assumption that this trend is explained mainly by density and polarizability parameters. The tenability obtained is reversible, and upon washing the work function returns to its original value. The change in the work function between the lowest ($Si/SiO_2$-$PrNH_2$) and the highest work function ($Si/SiO_2$-$PrNH_2$ with HI) was found to be of 0.86 eV – almost the whole band gap of Si. This might prove useful in biosensors applications for probing ion dynamics.

## Acknowledgments

This investigation was promoted with financial support of the H2020-FETOPEN project Reservoir Computing with Real-time Data for future IT (RECORD-IT) under grant nr. 664786; SY is the Binjamin H. Birstein Chair in Chemistry. TU Dresden gratefully acknowledged computing time by the Center

**TOC**

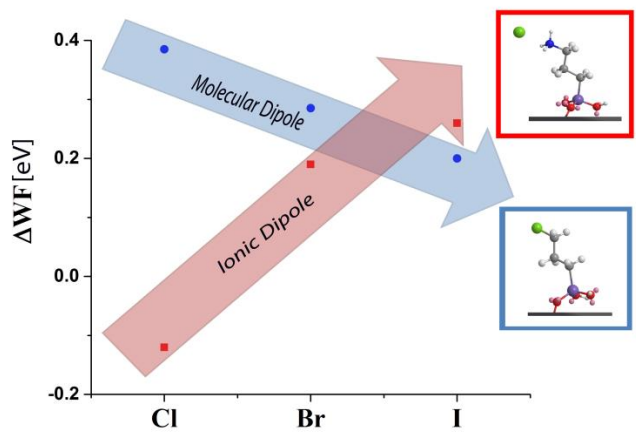